# Magnetic Clouds: Solar Cycle Dependence, Sources, and Geomagnetic Impacts

Y. Li • J. G. Luhmann • B. J. Lynch

**Abstract.** Magnetic clouds (MCs) are transient magnetic structures giving the strongest southward magnetic field (Bz south) in the solar wind. The sheath regions of MCs may also carry southward magnetic field. Southward magnetic field is responsible for causing space-weather disturbances. We report a comprehensive analysis of MCs and Bz components in their sheath regions during 1995 to 2017. Eighty-five percent of 303 MCs contain a south Bz up to 50 nT. Sheath Bz during the 23 years may reach as high as 40 nT. The MCs of strongest magnetic magnitude and Bz south occur in the declining phase of the solar cycle. The bipolar MCs have solar-cycle dependence in their polarity, but not in the occurrence frequency. Unipolar MCs show solar-cycle dependence in their occurrence frequency but not in their polarity. MCs with the highest speeds, largest total B magnitudes and sheath Bz south are from source regions closer to the solar disk center. About 80% of large Dst storms are caused by MC events. The combinations of south Bz in the sheath and the south-first MCs in close succession have given the largest storms. The solar-cycle dependence of bipolar MCs is extended to 2017, spanning 42 years. We find that the bipolar MC Bz polarity solar-cycle dependence is given by MCs originated from quiescent filaments in decayed active regions and a group of weak MCs of unclear sources, while the polarity of bipolar MCs with active-region flares always has mixed Bz polarity without solar-cycle dependence and is therefore the least predictable for Bz forecasting.

**Key words.** Coronal Mass Ejection • Magnetic Clouds

Y. Li
yanli@ssl.berkeley.edu

Space Sciences Laboratory
University of California at Berkeley
7 Gauss Way
Berkeley, CA 94804, USA

## 1. Introduction

Coronal mass ejections (CMEs) may carry enhanced magnetic field and fast-moving plasma into the heliosphere. The magnetic field of a CME originates in the low solar atmosphere, where a stressed magnetic structure becomes unstable and erupts, escaping the Sun. The interplanetary CMEs (ICMEs) that exhibit the *in-situ* topology of helical magnetic flux ropes in the interplanetary magnetic field (IMF) observations are referred to as magnetic clouds (MCs) (e.g. Burlaga, 1982; Bothmer and Schwenn, 1998). Herein, we will use the Mulligan, Russell, and Luhmann (1998) classifications that describe the behavior of the Bz component within MC flux ropes corresponding to axis orientations with low inclination and high inclination with respect to the ecliptic plane as bipolar and unipolar MCs, respectively. When a bipolar MC passes the observer, the IMF time series often shows a large component (and rotation) of Bz within the MC, which can be either S (Southward) in the leading portion and N (Northward) in the trailing portion, defined as a SN MC, or the reverse, defined as a NS MC. Unipolar MCs generally have only a S or N Bz component, referred to as S MCs or N MCs, respectively. Thus the magnetic field of MCs may present four different types: SN, NS, S and N. MCs often carry long-lasting steady southward Bz field, making them important for space weather concerns (*e.g.* Zhang *et al.*, 2004; Zhang *et al.*, 2007; Gopalswamy *et al.*, 2008; Richardson and Cane, 2013). The prediction of Bz south field has been a challenging task and attracted coordinated efforts (Riley, 2016; Riley and Love, 2017). The Bz component of the interplanetary field plays an important role in the coupling of the solar wind and IMF with the magnetosphere of the Earth. MCs also compress and disturb the ambient solar wind and IMF to form sheath regions ahead. Fast MCs often drive shocks, and the sheath regions between the shock and the magnetic flux rope can have an enhanced Bz component.

The Bz fields within MCs have been studied by many, while Bz fields in their sheath regions have not been systematically studied. We obtain the sheath southward Bz for 23 years starting 1995 in this article. The Bz polarity in bipolar MCs has been found to show a dependence on the solar cycle (Zhang and Burlaga, 1988; Bothmer and Rust, 1997; Bothmer and Schwenn, 1998; Mulligan, Russell and Luhmann, 1998). Li *et al.* (2011, 2014) extended these results, showing that Bz in bipolar MCs has a cyclic reversal on the time scale of the solar magnetic cycle over three sunspot cycles. It has been reported that about a third of ICMEs are MCs (Gosling, 1990). The ratio between MCs and ICMEs are solar-cycle dependent, and almost all ICMEs are MCs near solar minimum, but the proportion of MC ICMEs near solar maximum is much lower (Richardson and Cane, 2010). During Solar Cycle 23, approximately 48% of ICMEs are MCs, and 40% of MCs are bipolar MCs (Li *et al.*, 2011, 2014). The speed of the bipolar MCs has essentially the same distribution as all ICMEs, which implies that they are not from any special group of CMEs in terms of the solar origin. Although CME flux ropes may undergo a number of changes during the eruption and propagation processes or be sampled *in-situ* with a variety of impact parameters, a significant number of MCs evidently retain sufficient similarity to the orientation of their source-region magnetic field to possess the same cyclic periodicity in polarity reversal. During the solar minima, the Bz field at the leading portion of a bipolar MC is the same as the solar global dipole field (also noted in Mulligan, Russell, and Luhmann, 1998). This finding suggests that MCs preferentially remove the like polarity of the solar dipole field, thereby supporting the idea that CMEs play a role in the solar magnetic cycle. The solar-cycle dependence of MC orientation is interesting not only because it shows the connection between the MC magnetic field and the solar magnetic field but also because, as mentioned above, the large and long-lasting Bz field component has important space weather implications.

Since 2007, the twin *Solar Terrestrial Relations Observatory* (STEREO) spacecraft have been orbiting the Sun at about 1 AU near the ecliptic plane in addition to the *Advanced Composition Explorer* (ACE) and *Wind* spacecraft at the $L_1$ point, making three independent observing points of the *in-situ* solar wind and IMF near the ecliptic plane at 1 AU during an extended period for the first time. Moreover, STEREO images offer stereo views of the corona and CMEs, which are extremely useful in accurately identifying the source regions of CMEs.

In this article, after a brief description of the data source and events selection (Section 2), we present a comprehensive study of MCs encountered at the $L_1$ point during the years 1995 to 2017, when we have continuous high quality data. We analyze both bipolar and unipolar MCs to investigate the solar-cycle dependence of the Bz field, as well as the south Bz component in the sheath region ahead of the MCs (Section 3). We then identify the solar sources for MCs at $L_1$ during years 2007 to 2017, when multiple views of the Sun are realized after the launch of STEREO. We separate bipolar MCs into groups by the types of their source regions and examine the solar-cycle dependence of Bz field for each group (Section 4.1). We also identify and divide the solar sources into groups by the types of MCs they give rise to and present their locations in solar coordinate and against the background of the magnetic butterfly diagram (Section 4.2). We then study the association of the MCs with geomagnetic storms measured by the Dst index (Section 5). Finally, we update the MC polarity solar-cycle dependence by extending the study through 2017 using bipolar MCs encountered both at the $L_1$ point and the STEREO twin spacecraft sites (Section 6). We conclude the paper with discussions and conclusions regarding the overall MC characteristics and their implications (Section 7).

## 2. MC Event Selections

We use solar wind plasma and IMF *in-situ* measurements on ACE at the $L_1$ point and on the STEREO twin spacecraft orbiting the Sun at ≈1AU (treating the observations on each spacecraft as an independent data set). We analyze five-minute ACE/The Magnetic Experiment (MAG) and /The Solar Wind Electron, Proton, and Alpha Monitor (SWEAPAM) merged level 2 data, and STEREO-A and -B 10-minute In-situ Measurements of Particles and CME Transients (IMPACT) and Plasma and Suprathermal Ion Composition (PLASTIC) merged level 2 data. By examining the IMF three components and the solar-wind plasma parameters of proton temperature, density and bulk speed, and the plasma beta, we select MC intervals. To identify MCs, we require i) an enhanced magnetic field magnitude greater than ≈ 8nT, ii) a low-variance magnetic field with a coherent rotation of the field vector over a time interval on the order of a day, and iii) a lower-than-average proton temperature (*e.g.* Burlaga, 1988). The selection process is by visual inspection, in the same manner as previous studies for consistency (Mulligan, Russell, and Luhmann, 1998; Li *et al.*, 2011, 2014), except that we lowered the requirement for magnetic enhancement to be no less than 10 nT to no less than 8 nT because some good MC events can be missed, especially in Solar Cycle 24 when the magnetic field is weaker (*e.g.* see Lee *et al.*, 2009; Kilpua *et al.*, 2014).

Note that in this study we do not include structures that can be classified as ICMEs but do not have a clearly identifiable MC driver (*e.g.* see Jian *et al.*, 2006). As discussed by Jian *et al.* (2006) and other authors, the general consensus is that MCs are generally present in ICMEs as drivers but are best observed when a spacecraft/observer is more centrally located within the passing structure. Thus some of our results apply mainly to the more centrally sampled ICMEs. Also, the initial flux ropes in CMEs can be distorted during their propagation, *e.g.* due to interactions between CMEs (Lugaz *et al.*, 2017).

With these criteria, we have identified 303 MCs at $L_1$ using ACE and/or OMNI data (https://omniweb.gsfc.nasa.gov/ow.html) during the years 1995 to 2017, including 194 bipolar MCs and 109 unipolar MCs. All of the MCs at $L_1$ point and their sheath southward Bz field, as well as their source regions, will be studied and presented in the next section. We have identified 67 bipolar MCs at STEREO-A and -B during the years 2007 to 2013 due to data availability and continuity issues in and after 2014. We have also identified eight bipolar MCs at STEREO-A in the years 2016 and 2017. Bipolar MCs from all three spacecraft are used to extend the record of their polarity, long-term variation, and dependence on the solar cycle at the end of the article.

## 3. Magnetic Field in MCs and Sheath Regions

Since 1995, the *in-situ* data at $L_1$ have good quality and essentially continuous coverage. We found 194 bipolar MCs and 109 unipolar MCs during the 23 years from 1995 to 2017, including 103 SN, 91 NS, 64 S, and 45 N MCs. We note that there are only 45 north-only (N type) MCs out of the total 303 MCs, which means 258 MCs contain south Bz in their internal field, *i.e.* 85% of MCs contain a south Bz component in part or for the entire duration of the MC passage over the Earth, clearly showing the importance of MCs for geo-disturbances.

Figure 1a presents the annual counts of NS and SN bipolar MCs versus time, Figure 1b has the annual counts of the S and N unipolar MCs versus time, Figure 1c shows the annual counts of bipolar MCs as positive values and unipolar MCs as negative values, Figure 1d gives the annual counts of total MCs in open histogram and those MCs with a south Bz component in filled histogram, Figure 1e shows the normalized polarity histogram based on the annual counts in Figure 1a, and Figure 1f shows the sunspot numbers of Cycles 23 and 24. The vertical lines in light gray mark the two solar minima in the years 1996 and 2009, and in dark-gray lines mark the two solar maxima in the years 2000 and 2014. The polarity of the bipolar MCs has a clear solar-cycle dependence, as known previously. The polarity of the unipolar MCs, however, has no such trend. The occurrence of unipolar MCs is much less frequent around solar minima. The number of bipolar MCs is greater than that of unipolar MCs as a whole and also in each year except 2001. The occurrence of MCs (Figure 1d) does not have a clear solar-cycle dependence while the occurrence of general ICMEs does, as reported previously (Jian *et al*., 2006; Jian, Russell, and Luhmann, 2011; Richardson and Cane, 2010). Note this is only concerned with the occurrences; the south magnetic-field magnitude within MCs will be shown next.

In addition to the plasma and magnetic field properties within MCs, the magnetic-field strength and polarity in the ICME sheath regions ahead of MC ejecta are also important parameters in terms of the space-weather effects of these events. In Figure 2, we present these parameters *versus* time: (a) maximum ICME plasma speed, (b) maximum magnetic-field magnitude within MCs, (c) maximum value of Bz south within MCs, (d) maximum value of Bz south in the MC sheath region, and (e) the sunspot numbers from 1995 to 2016 spanning Cycles 23 and 24. Again, the light- and dark-gray lines mark the solar minimum and maximum, respectively. The red and blue circles represent parameters for SN and NS bipolar MCs, respectively. The black and green plus symbols are for S and N unipolar MCs, respectively. The red, blue, black, and greed squares represent parameters of the sheath regions ahead of SN, NS, S, and N MCs, respectively. In Solar Cycle 23 (24), the MC speeds range 300 to 1000 (700) km s$^{-1}$, the magnetic magnitudes within MCs range 8 to 62 (40) nT, the south magnetic fields within MCs (*Bs*) range 0 to 45 (22) nT, and the south magnetic fields in the MC sheaths range 0 to 42 (17) nT. The ICMEs are slower and field strengths are much weaker in Cycle 24. In Cycle 23, the largest values of all parameters appear in the declining phase of the solar cycle between the solar maximum and the next solar minimum. Cycle 24 seems to have the same tendency, but it is less obvious, perhaps for two reasons; first that the values are all less significant, and second that the cycle is still not complete.

The south magnetic field within MCs (*Bs*) and in the sheath regions (***B****ss*) are reported here for the two solar cycles, while we have previously shown corresponding results for speed and total magnetic magnitude of MC in Li *et al*. (2011). From Figure 2, MC speed (*V*), total magnetic field magnitude (*B*, *Bs*, and ***B****ss*) are all solar-cycle dependent, being weaker around solar-activity minimum and stronger around solar-activity maximum and during a large part of the declining phase. The geoeffectiveness of the MC events and the combined effect of strong *Bs* and ***B****ss* will be discussed in Section 7.

In Figure 3, we present MCs analyzed above versus ICMEs found by a few other studies. Figure 3a presents the MCs as a black line and ICMEs from three other research groups in gray lines of different shades (http://www.srl.caltech.edu/ACE/ASC/DATA/level3/icmetable2.htm, Richardson and Cane, 2010; http://space.ustc.edu.cn/dreams/wind_icmes/, Chi *et al*., 2016; http://www-ssc.igpp.ucla.edu/~jlan/ACE/Level3/ICME_List_from_Lan_Jian.pdf, Jian *et al*., 2006, 2011, 2013), Figure 3b has the ratio of MCs versus ICMEs, and Figure 3c shows the sunspot numbers spanning Cycles 23 and 24 again as reference. The vertical lines in light-gray mark the two solar minima in 1996 and 2009, and those in dark-gray mark the two solar maxima in 2000 and 2014. The occurrence of MCs (Figure 3a) does not have a solar-cycle dependence while the occurrence of ICMEs does, as previously reported (Jian *et al*., 2006, 2011; Richardson and Cane, 2010). The ratio between MCs and ICMEs (Figure 3b) has a solar-cycle dependence with the maximum ratio as high as unity at solar minima, and as low as 0.3 at some point on the rising phase of the solar activity (using sunspot numbers in Figure 3c for reference).

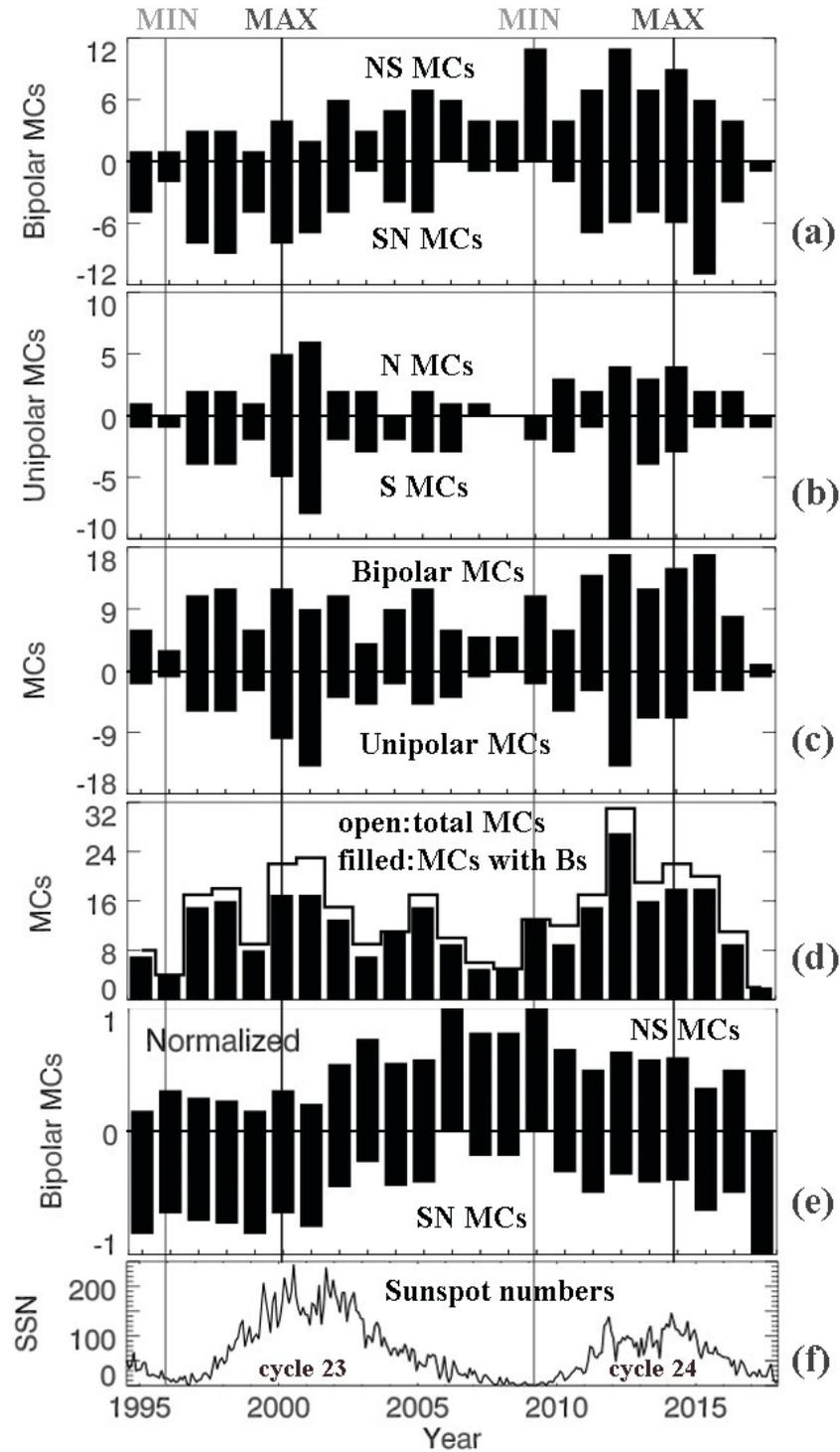

**Figure 1.** (**a**) Annual counts of bipolar MC NS and SN MCs; (**b**) annual counts of unipolar N and S MCs; (**c**) annual counts of bipolar and unipolar MCs; (**d**) annual counts of MCs (*open bars*) and those MC containing south Bz fields (*filled bars*); (**e**) normalized polarity histogram based on the annual counts in Figure 1a; (**f**) sunspot numbers of Solar Cycles 23 and 24.

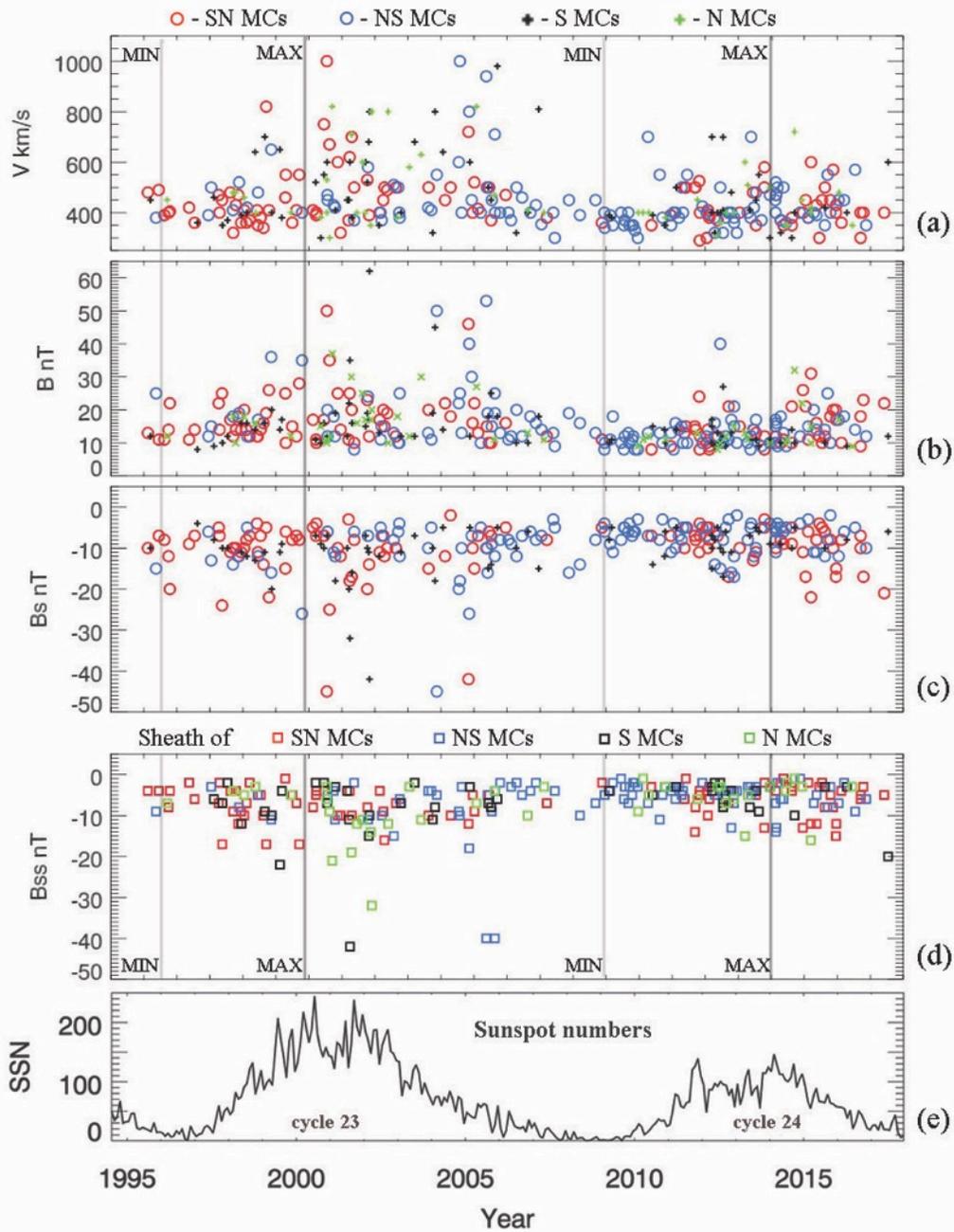

**Figure 2.** (**a**) MC plasma speed; (**b**) peak value of magnetic-field magnitude within MCs; (**c**) peak value of Bz south within MCs; (**d**) peak value of Bz south in the sheath region before MCs; and (**e**) the sunspot numbers for reference. The *light-* and *dark-gray lines* mark the solar minimum and maximum, respectively. The *red and blue circles* represent parameters of SN and NS MCs, respectively. The *black and green plus sign* are for S and N MCs, respectively. The *red, blue, black and greed squares* represent parameters of the sheath regions before SN, NS, S, and N MCs, respectively.

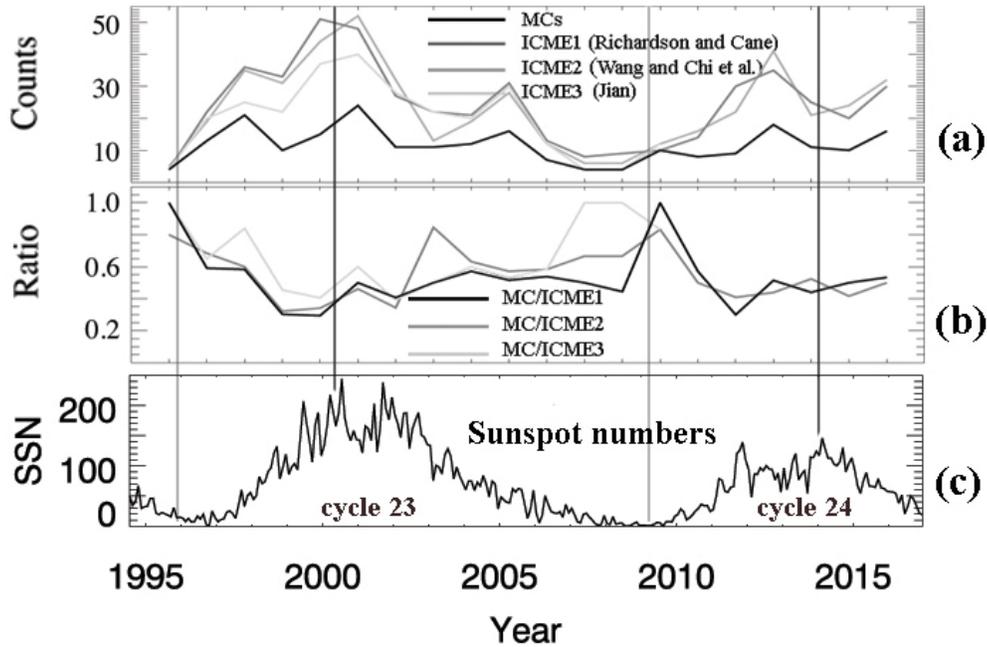

**Figure 3.** (**a**) time series of MC annual counts and ICME annual counts adapted from three other authors; (**b**) ratio of each year between MCs and ICMEs; (**c**) sunspot numbers of Solar Cycles 23 and 24.

**4. MCs and Solar Sources**

Since 2007, when the STEREO twin spacecraft launched, we have had multi-perspective views of the Sun, as well as full-disk coronal images in higher cadence and with better availability than had previously been available. In 2010, the launch of the *Solar Dynamics Observatory* (SDO) provided the community with Earth-view full disk images in higher cadence and resolution, and excellent quality. With these unprecedented resources, researchers are able to more confidently make the associations between *in-situ* MC events and their parent CMEs, including identifying the source regions of the CMEs in the lower solar atmosphere. In our identification of the solar sources of the $L_1$ MCs, we utilized the images and movies from the Large Angle Spectrometric Coronagraph (LASCO) CME catalog and STEREO Science Center, the SDO images and movies from the Sun in Time website (http://sdowww.lmsal.com/suntoday_v2/index.html) made available by Lockheed-Martin Solar and Astrophysics Lab (LMSAL), and the *Geostationary Operational Environment Satellite* (GOES) soft X-ray flare and other information from the Solar Monitor website (https://www.solarmonitor.org). We first search the LASCO CME catalog for halo, partial halo or wide CMEs greater than 60° during the five days prior to the MC arrival. Often, there are multiple candidate sources for a CME during active periods of the Sun near solar maximum. In these cases, multi-perspective observations made possible with the STEREO twin spacecraft are particularly helpful. Multi-views are also extremely useful to identify backside halo CMEs. STEREO coronagraph and Heliospheric Imagers (HI) images are examined whenever available for better certainty of the correspondence between a CME and the MC at $L_1$ point. We further identify the associated activity in the low corona, including solar flares, filament eruptions, and EUV dimmings (*e.g.* Thompson *et al.*, 1998; Krista and Reinard, 2017). Clearly, there is an added step to identify sources for MCs than CMEs, and therefore more uncertainty involved.

During 2007 to 2017, 149 MCs including 100 bipolar MCs and 49 unipolar MCs have been found that encountered the Earth. 42 MCs had active-region flare sources and 57 MCs had filament eruption sources. For 8 MCs, only dimming was seen in EUV movies as the activity source and one MC had a weak eruption-like signature in EUV movies as the source. 40 MCs had unclear sources. When an MC event is marked with a flare source, for example a GOES soft X-ray flare is recorded greater than ≈B8.0 from an active region, we take the flare over the other CME signatures present for the classification of the MC source. When an MC is marked with a filament-eruption source, the filament is a large one in a decayed active region. A large filament eruption sometimes may also have a soft X-ray flare, in this case we take the filament over the flare for the classification of the MC source. When an MC is identified with a dimming source, the dimming in the EUV images or movies is the only signature associated with the CME event. For the MCs with no clear sources include these cases: i) identified parent CME has no identifiable eruption signature at low corona (11 out of 40 MCs); ii) no suitable coronagraph CME in LASCO CME catalog in our five-day window (13 out of 40 MCs); iii) uncertain parent CMEs: multiple possible source CMEs (eight out of 40 MCs) and iv) uncertain sources: multiple possible source activities (ten out of 40 MCs). The first case of 11 CMEs corresponds to the so-called "stealth" CMEs (Robbrecht *et al.*, 2009; D'Huys *et al.*, 2014; Alzate and Morgan, 2017).

**4.1 Bipolar MCs Grouped by Solar Source Type**

Among the 100 bipolar MCs, 27 had flare sources, 39 had filament sources, 27 had unclear sources, 6 had dimming sources, and 1 had a small eruption-like source. We include the first three types in Figure 4 as three separate bar plots to investigate their polarity variation with time. Figure 4 shows the polarity of bipolar MCs from (a) flare sources, (b) filament sources, and (c) unclear sources. Figure 4d shows sunspot numbers over the same time range. The flare-source MCs rarely occur in the period of low solar activity around solar minimum, but they occur much more frequently around solar maximum, as expected. No solar-cycle polarity trend exists for this group of bipolar MCs with flare sources, unlike what was described in the previous section for the set of all bipolar MCs. The bipolar MCs with filament sources do reflect the solar-cycle polarity trend. If one ignores the one SN outlier MC in 2011, the solar-cycle trend is better shown. Considering that the number of events is so small, we conclude that this group does show the same solar-cycle dependence with fluctuations. The MCs with unclear sources may also be considered as having the same solar-cycle dependence. In fact, the NS MCs around the last solar minimum were mostly from the contributions of the third group of weak MCs with unclear sources including stealth CME. Since stealth CMEs are from the coronal streamers (*e.g.* see Lynch *et al.*, 2016, and references therein), these MCs' field directions might be expected to agree with the solar dipole field.

By separating MCs into the three groups, only now do we have a clear picture that the bipolar MC polarity solar-cycle dependence is given by those MCs originated from quiescent filaments in decayed active regions and the group of weak MCs of unclear sources including stealth CMEs.

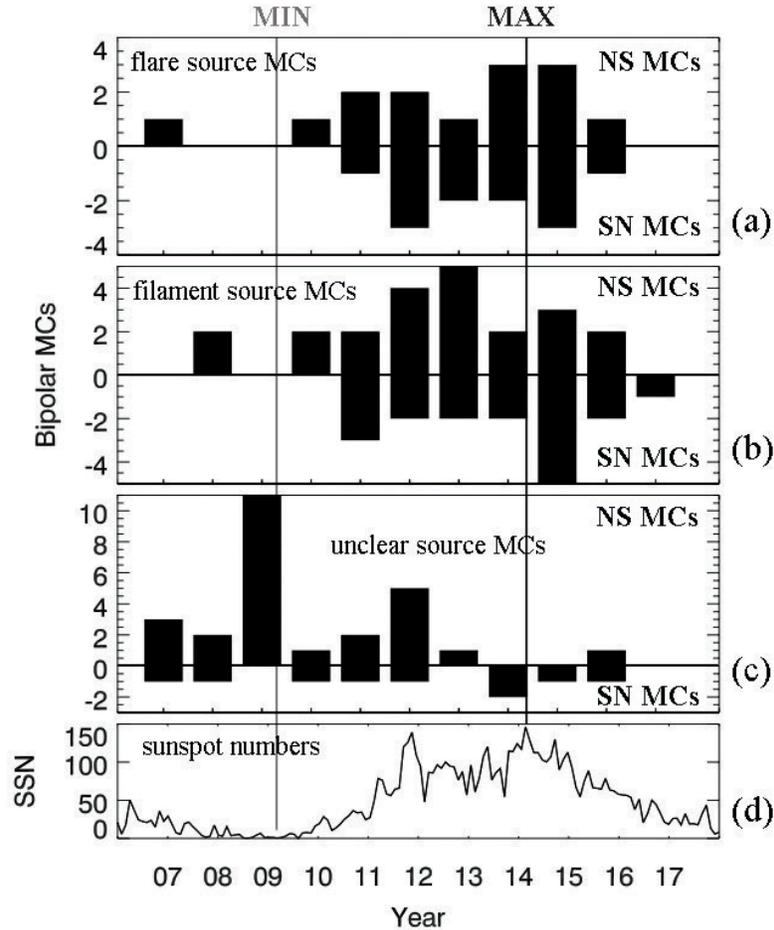

**Figure 4.** Bipolar MCs separated in to groups according to different types of sources. Annual counts of NS and SN MCs originated from active region flares (**a**); large quiescent filaments (**b**); and unclear sources (**c**). The bottom panel shows the sunspot numbers of year 2007 to 2017 (**d**).

### 4.2 MC Solar Sources and Parameters

For MCs (including both bipolar and unipolar) having identifiable sources, 42 MCs had flare sources and 57 MCs had filament eruption sources. For eight MCs, only dimming signatures are observed in EUV images. We recorded the locations of these source regions, and present them in Figure 5 in solar coordinates as seen from Earth, and in Figure 6 in synoptic format over the magnetic butterfly diagram up to the year 2016 by D. Hathaway at NASA (https://solarscience.msfc.nasa.gov/dynamo.shtml). In Figures 5 and 6, red symbols represent flares, blue symbols represent filaments, green symbols are for dimming cases, solid symbols are for north-first or north-only MCs and open symbols are for south-first or south-only MCs; for further details see figure annotations and captions. Figure 5a includes all events, Figure 5b includes bipolar MCs and Figure 5c includes unipolar MCs. Most of the sources are located within 45° latitudes and longitudes with respect to the subsolar point (highlighted by orange lines), except for a few outliers. Flare sources in general are located at lower latitudes than large filament sources, which is reasonable considering the filament association with decayed active regions. More source regions are located in the western hemisphere than the eastern hemisphere (Hess and Zhang, 2017). The source regions do not appear to be distinguished or ordered by their resulting MC types in this format of display.

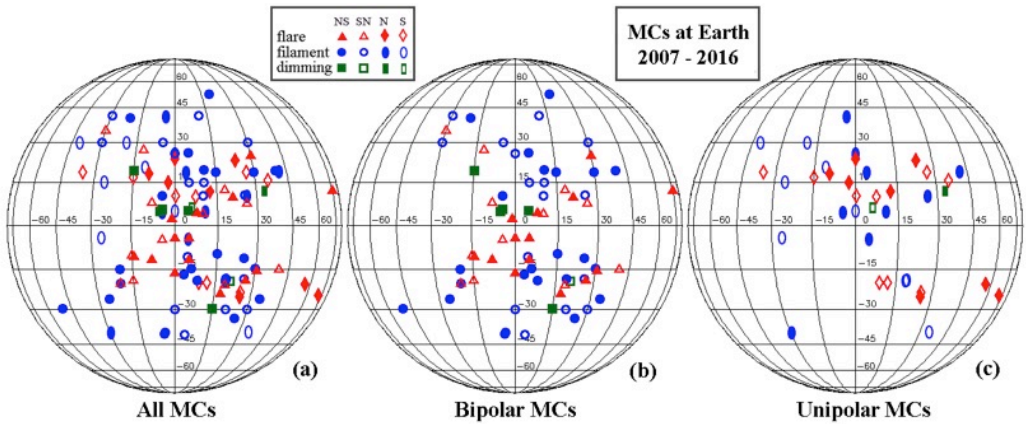

**Figure 5.** The locations of all types of MC source regions in the solar coordinate system. Specifically, (**a**) source regions for all MCs in years 2007 to 2016 at $L_1$; (**b**) source regions of bipolar MCs; (**c**) source regions of unipolar MCs. The symbols: *filled-red triangles* are for flares giving rise to NS MCs; *open-red triangles* are for flares giving rise to SN MCs; *filled-red diamonds* are flares for N MCs; *open-red diamonds* are flares for S MCs; *filled-blue circles* are filaments for NS MCs; *open-blue circles* are filaments for SN MCs; *filled-blue ellipses* are filaments for N MCs; *open-blue ellipses* are filaments for S MCs; and *filled-green squares* are EUV dimming for NS MCs.

Figure 6 includes the solar source locations of (a) all events, (b) bipolar events, and (c) unipolar events. From 2007 to about 2014, more source regions are seen in the northern hemisphere, but after 2014 more sources are seen in the southern hemisphere (Hess and Zhang, 2017). In Figures 6d, 6e, and 6f, we presented the bipolar MC subset of solar source locations for flare sources, filament sources, and dimming sources, respectively. The flare sources show the most agreement with the butterfly patterns formed by active regions, as expected. This format shows more clearly that the large filament sources are located at higher latitudes than flares sources as a whole. In Figure 6d, flare sources for NS MCs (filled-red triangles) and SN MCs (open-red triangles) are quite mixed and not ordered by time as shown in Figure 4a in bar-plot form. In Figure 6e, filament sources for NS MCs (filled-blue circles) and SN MCs (open-blue circles) are separated. Here, the mostly filled-blue circles starting 2007 switch to being mostly open-blue circles towards 2017 as shown in Figure 4b in a bar plot. In Figure 6f, there are too few cases to show any well-defined trend, although filled-green squares (sources for NS MCs) occur first, with the last one in 2015 giving way to open-green squares (sources for SN MCs). These MC sources with only dimming signatures on the solar disk may correspond to stealth CMEs when the sources are on the disk instead of at the limb. If this is a correct interpretation, these events may be added to the group shown in Figure 4d, where the bar plot indicates that the polarities of the resulting bipolar MCs have clear solar-cycle dependence.

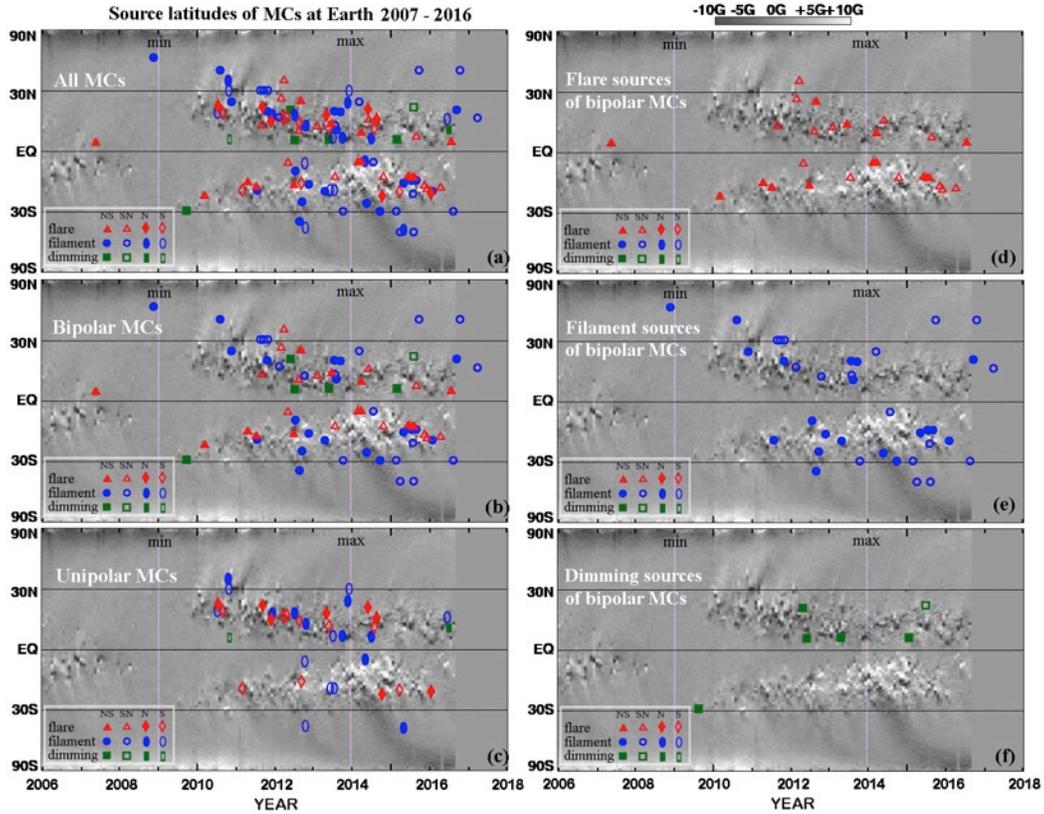

**Figure 6.** The locations of MC source regions overlaid on the magnetic butterfly diagram by D. Hathaway (NASA). The magenta lines mark solar minimum and maximum. (**a**) source regions for all MCs in years 2007 to 2017 at $L_1$ on butterfly diagram; (**b**) source regions of bipolar MCs on butterfly diagram; (**c**) source regions of unipolar MCs on butterfly diagram; (**d**) flare sources of bipolar MCs; (**e**) large filament sources of bipolar MCs; and (**f**) dimming sources of bipolar MCs. The symbols are the same as in Figure 5. The symbols: *filled-red triangles* are for flares giving rise to NS MCs; *open-red triangles* are for flares giving rise to SN MCs; *filled-red diamonds* are flares for N MCs; *open-red diamonds* are flares for S MCs; *filled-blue circles* are filaments for NS MCs; *open-blue circles* are filaments for SN MCs; *filled-blue ellipses* are filaments for N MCs; *open-blue ellipses* are filaments for S MCs; and *filled-green squares* are EUV dimming for NS MCs.

Next, we investigate whether there is any association between MC parameters at the Earth and their solar source location. The MC speed and magnetic-field strength, particularly the B south, are of the most space-weather concerns. In Figure 7, we present four scatter plots of MC parameters with normalized distance to solar disk center [$r/R$], where $r$ is the distance of a MC source to solar disk center and defined using source latitude [$\theta$] and longitude [$\phi$] in solar coordinates:

$$r/R = \sqrt{\cos^2\theta \sin^2\varphi + \sin^2\theta}$$

For MCs of unclear sources, we assign $r/R = 1.2$ arbitrarily for displaying. Figure 7a shows the scatter of MC maximum speed V *versus r/R* in black-triangle symbols, and the vertically lined up gray triangles are the speed of MCs of unclear sources, which are all less than around 500 km s$^{-1}$. It appears that all MCs faster than 600 km s$^{-1}$ have sources within $r/R < 0.6$ except for one outlier. Figure 7b shows the scatter of MC maximum magnetic-field magnitude [$B$] *versus r/R* in black plus symbols, and the gray plus symbols are for MCs of unclear sources. MCs of maximum $B$ magnitude stronger than 25 nT appear to have sources within $r/R < 0.6$ without exceptions. MCs of unclear sources all have maximum $B$ magnitude below 20 nT. Figure 7c shows the scatter of MC maximum southward magnetic field $B_s$ *versus r/R* in black-cross symbols, and the gray-cross symbols are for MCs of unclear sources. $B_s$ appears to be less ordered by the MCs' source location than $B$ magnitude, but it does also show a weak tendency of having stronger values with sources closer to solar disk center. Only one MC has $B_s$ stronger than -20 nT in Solar Cycle 24 by the end of year 2016. Figure 7d shows the scatter of southward magnetic field in MC sheath [$B_{ss}$] *versus r/R* in black-diamond symbols, and the gray diamonds are for MCs of unclear sources. $B_{ss}$ shows a noticeable trend as being stronger when MC sources closer to the solar disk center. Values of $B_{ss}$ for MCs with unclear sources are all below 8 nT with one exception.

## 5. Geomagnetic Impacts of Magnetic Clouds

Geomagnetic storms can be measured by the Dst index, and larger depressions of the Dst index indicate stronger storms. The Dst index is particularly sensitive to the IMF south Bz field and the solar wind dynamic pressure (*e.g.* Kilpua *et al.*, 2014). In Solar Cycle 23 (1996 to 2008), there were 17 very large storms of Dst < -200 nT. We used (hourly) Dst values in Omni2 data at the OMNIWeb NASA GSFC. Out of the 17 storms, 14 storms (82.4%) were caused by MC events. Out of these 14 storms, the corresponding MCs and their sheath regions both had south Bz in ten events, one event had south Bz in only the MCs and three events had south Bz in only the sheath regions. The current Solar Cycle 24 (2009 to 2017) had 19 large storms of Dst < -100 nT (including only two very large geomagnetic storms of Dst < -200 nT). Out of the 19 large storms, 14 storms (73.7%) are caused by MC events. Out of the 14 storms, the corresponding MCs and their sheath regions both have south Bz in 11 events, two events have south Bz in only the MCs and one event has south Bz in only the sheath region.

The three largest geomagnetic storms in Cycle 24 have minimum Dst values of -220 nT, -201 nT, and -165nT. The sources of the largest and the third largest storms are MCs, which we will show as examples. Figure 8 shows the solar wind, IMF observations, and the associated Dst response for the two MC-driven storm intervals in Cycle 24. In Figure 8a, we see the MC (shaded in magenta) starting 17 March 2015 and the corresponding Dst time series for the largest storm. This is an SN MC; the front part of the MC has a smooth, southward Bz field lasting ≈15 hours with the minimum at ≈ -22 nT, and the MC speed is ≈ 600 km s⁻¹. This MC has the strongest B south in the current cycle, but is also moderately fast. The combination of the $B_s$ in the MC and the $B_{ss}$ in its sheath region (shaded in pale yellow) right before it caused the largest geomagnetic storm of the current solar cycle. The Dst in the bottom row shows that the initial depression corresponds to the sheath region ahead of the MC, where a south Bz field of minimum ≈ -12 nT is present, at the same time the solar-wind density in the sheath is much higher than that within the MC by a factor ≈ 5, which produces a larger dynamic pressure. The MC sheath caused the first stage of the storm, producing a Dst ≈ -70 nT, and the MC continues to cause the second stage of the storm. In Figure 8b, we have an MC (shaded in magenta) starting 20 December 2015 and the corresponding Dst time series for the third largest storm. This is also an SN MC, the south Bz in the MC lasted ≈ 24 hours with a minimum of ≈ -17 nT, and the speed is ≈ 400 km s⁻¹. The sheath region (shaded in pale yellow) ahead of the MC has south Bz of ≈ -15 nT at the peak, a higher speed of ≈ 500 km s⁻¹ and a solar wind density over ≈ 10 times greater than that within the MC. The corresponding Dst curve again shows that the combination of the sheath and MC driver caused this two-stage strong storm.

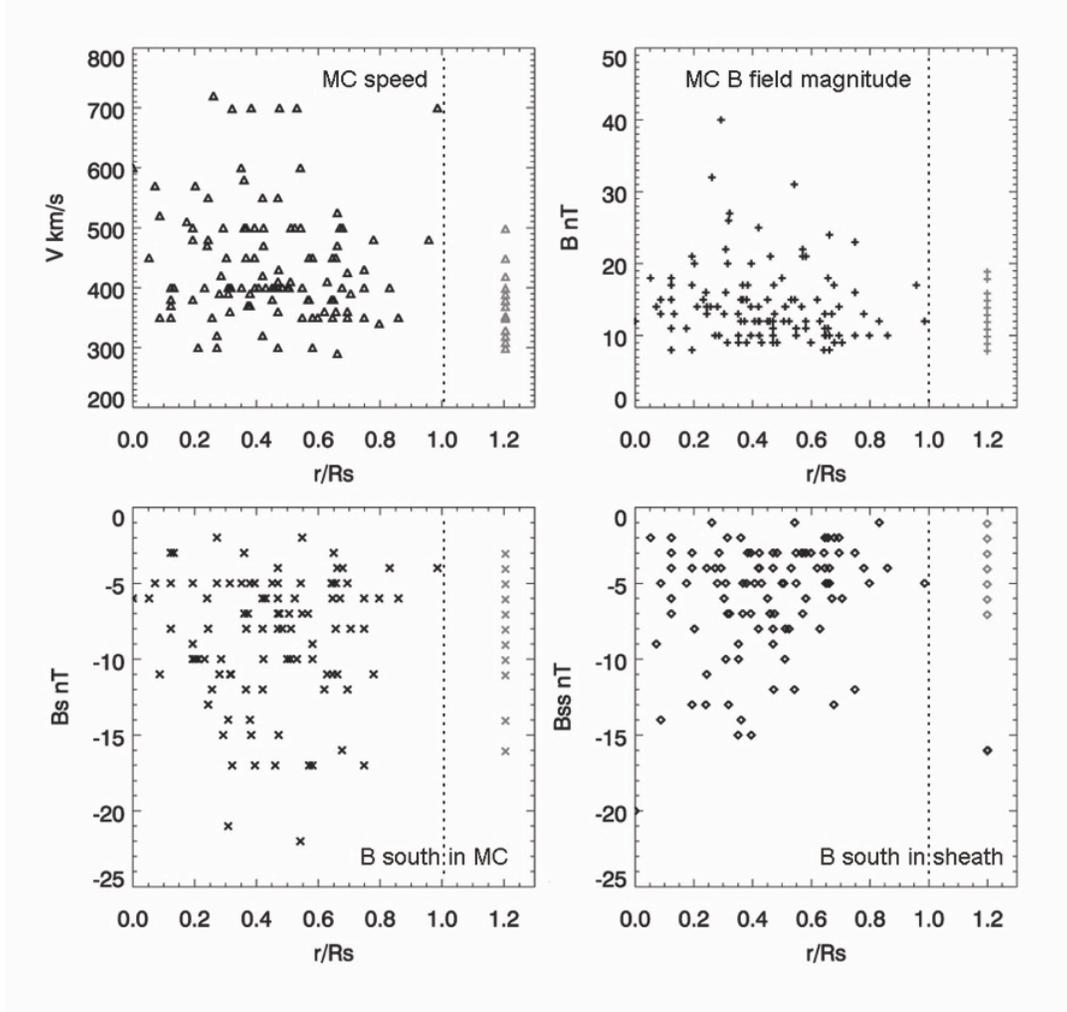

**Figure 7.** Scatter plots of MC parameters and normalized distance to solar disk center [r/R: see text for definition]. (**a**) maximum MC speed [V] vs. r/R; (**b**) maximum MC B field magnitude vs. r/R; (**c**) maximum MC B south [$B_s$] vs. r/R; and (**d**) maximum Sheath B south [$B_{ss}$] vs. r/R.

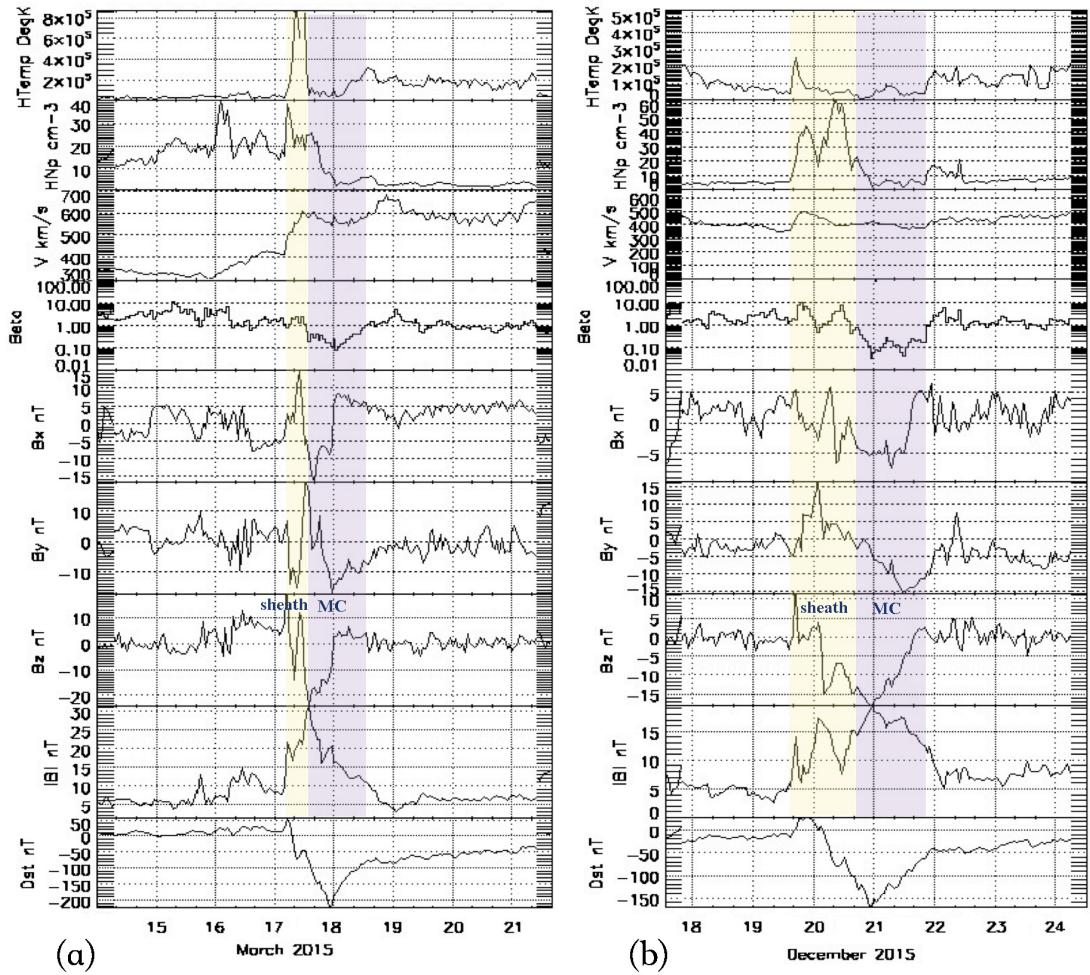

**Figure 8.** The solar wind and IMF measurements and Dst index of two time intervals of the strongest Dst magnetic storms. *From the top to bottom*, solar wind proton temperature, density, bulk speed, plasma β, IMF Bx, By, Bz and B magnitude; *the bottom tow shows the Dst index*. (**a**) the event in March 2015, (**b**) the event in December 2015, the *sheath regions and MCs are shaded in pale yellow and magenta*, respectively.

Both the south Bz in MCs and in their sheath regions are important for causing space-weather events and need to be taken into account in related studies and forecasting. During the current solar cycle, the majority of bipolar MCs have north-first field (NS MCs), while two of the three largest geomagnetic storms are due to two south-first MCs (SN MCs) combined with the southward field and large dynamic pressure in their sheath regions ahead of the MCs. Recall Figure 2 in Section 4, for the SN MCs in red-circle symbols and S MCs in black-cross symbols in Figure 2a. If south fields in their sheath regions [$B_{ss}$] as seen in the red- and black-square symbols in Figure 2d have large values, dual actions of the two parts of south B fields in close succession have good chances in causing large two-stage Dst storms. It is interesting to note that the green squares in Figure 2d are for Bz south fields in the sheath regions ahead of unipolar N MCs. In these cases, the sheath Bz south fields are the only geoeffective part of the events because the internal fields of N MCs produce little Dst response.

## 6. Long Term Solar Cycle Dependence of Bipolar MCs

It has been shown that, in bipolar MCs, the North-South component (Bz) reverses with the same periodicity as the solar magnetic field (Li *et al.*, 2011, 2014). Li *et al.* (2014) reported the solar-cycle dependence of the MC field polarity combining data from 1976 to 2012. In the current study, we extend the MC polarity long-term variation through the year 2016. The extended results, normalized as described below to better illustrate cycle phase relationships, are shown in Figure 9a (the previous study showed straight annual counts at each observing point). Here, a simple normalization has been applied to the number of NS (or SN) MCs as $f[yr]=n[yr]/n_t[yr]$, where $n$ is the number of NS (or SN) MCs in a year, and $n_t$ is the total number of the bipolar MCs of the same year. Figure 9 also provides the reference parameters of the solar cycle, including the sunspot numbers (b), the solar polar magnetic field (c); and the solar magnetic butterfly diagram up to year 2016 (d) made available by D. Hathaway at NASA. The gray bars in Figure 9a indicate where the data are sparse with frequent and long data gaps, and therefore the MC counts are not as reliable.

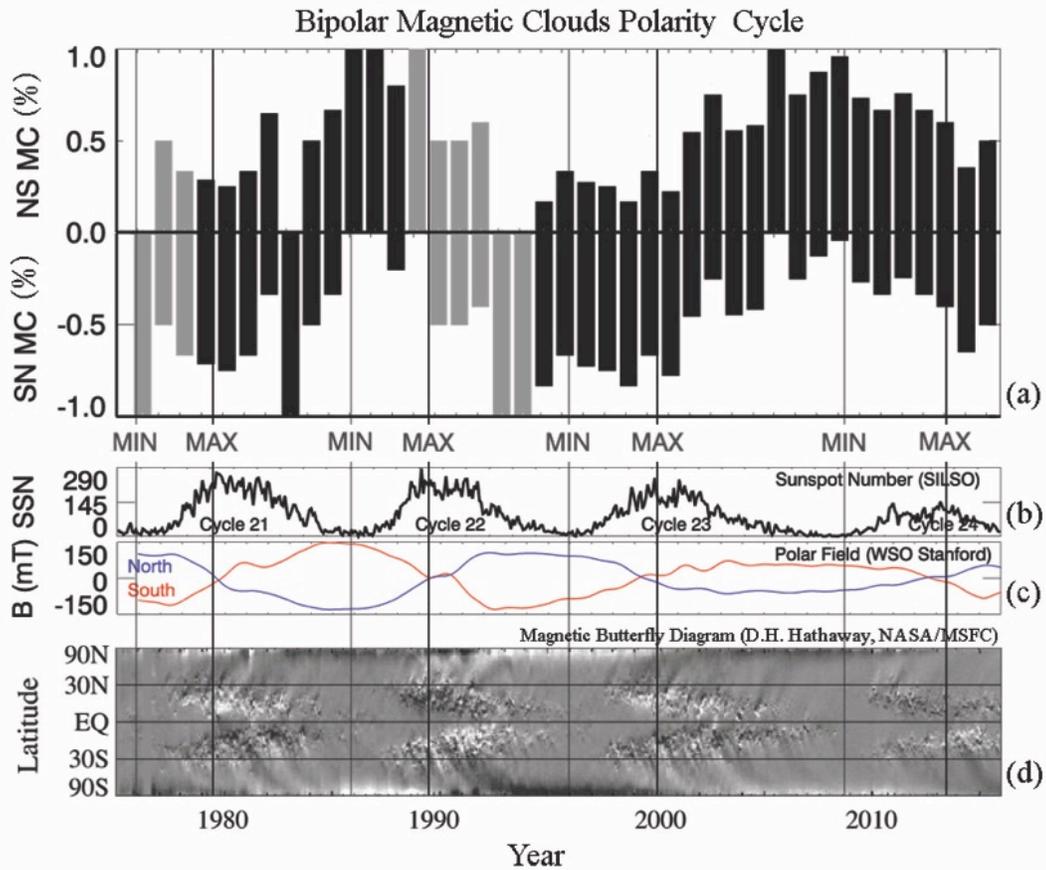

**Figure 9.** (**a**) Bipolar MC field solar-cycle dependence spanning four solar cycles, or 41 years, from 1976 to 2016. The gray bars indicate where the data are sparse with frequent and long data gaps, and therefore the MC counts are not as reliable. (**b**) sunspot numbers of four solar cycles; (**c**) solar polar magnetic field of four solar cycles; (**d**) solar magnetic butterfly diagram by D. Hathaway (NASA).

The MC polarity on the rising phase of the solar cycle between a solar minimum (MIN) and the next solar maximum (MAX) usually are clearly dominated by either NS or SN MCs. Between 1986 (MIN) and 1990 (MAX) MCs are mostly NS type, between 1996 (MIN) and 2000 (MAX) MCs are mostly SN type, and between 2009 (MIN) and 2014 (MAX) MCs are again back to mostly NS type. We note that during a couple of years around a solar minimum, MCs are always almost pure NS or SN with little exception. The dominant polarity reverses on the declining phase between the solar maximum and the next minimum. The last sunspot minimum and maximum occurred in 2009 and 2014, respectively. The Sun is currently on the declining phase of its activity cycle. The bipolar MC polarity in 2013 to 2017 shows large fluctuations between the numbers of the two types of bipolar MCs, while the overall polarity reversal of switching from majority NS MCs to having majority SN MCs is evident.

## 7. Discussions and Conclusions

Detailed analysis of modern era solar-wind and IMF data from 1995 to 2017 spanning the recent two Solar Cycles 23 and 24 have been carried out. We have studied both bipolar and unipolar MCs. In addition to their occurrences and orientation polarities, we investigated the solar-cycle and source-region dependence of the MC events' maximum speed, magnetic magnitude, Bz south component within the MCs, and in the sheath regions ahead of the MCs. The occurrence of bipolar MCs has less clear solar-cycle dependence, while their polarity does. Unipolar MCs mostly occur around solar active times, without solar-cycle dependence for their polarity. The solar-cycle dependence of the polarity of bipolar MCs has been known and generally accepted for decades. With the new data of recent solar cycles, we now make a quantitative analysis of this dependence and the dependence of MC occurrence on the solar cycle. Figure 10a, 10b, and 10c give the over plots of bipolar MC polarity with solar North (blue) and South (red) polar field values, and the bipolar and unipolar MC occurrences with sunspot numbers; and Figure 10d, 10e, and 10f show their scatter plots and the results of their linear fits. Figure 10d presents the bipolar MC polarity *vs.* the solar polar field strength, and the Pearson linear correlation coefficient [R] is ≈ 0.70, the Spearman rank correlation coefficients [ρ] give similar values of ≈ 0.70 with decent significance of p-value $< \approx 10^{-4}$. These analyses confirm a quite strong correlation of the polarity on the solar cycle. Figure 10e shows bipolar MC occurrence *vs.* sunspot numbers, and Pearson correlation is ≈ 0.4 and the rank correlation is 0.50 with much less significant of p-values of ≈ 0.01–0.02. The dependence of bipolar MC occurrence on solar cycle is much weaker, statistically. Figure 10f gives the unipolar MC occurrence *vs.* sunspot numbers, and Pearson correlation is ≈ 0.7 and the rank correlation is ≈ 0.7 with decent significance p-value $< \approx 10^{-4}$, similar to the polarity results and thus a clear solar-cycle dependence.

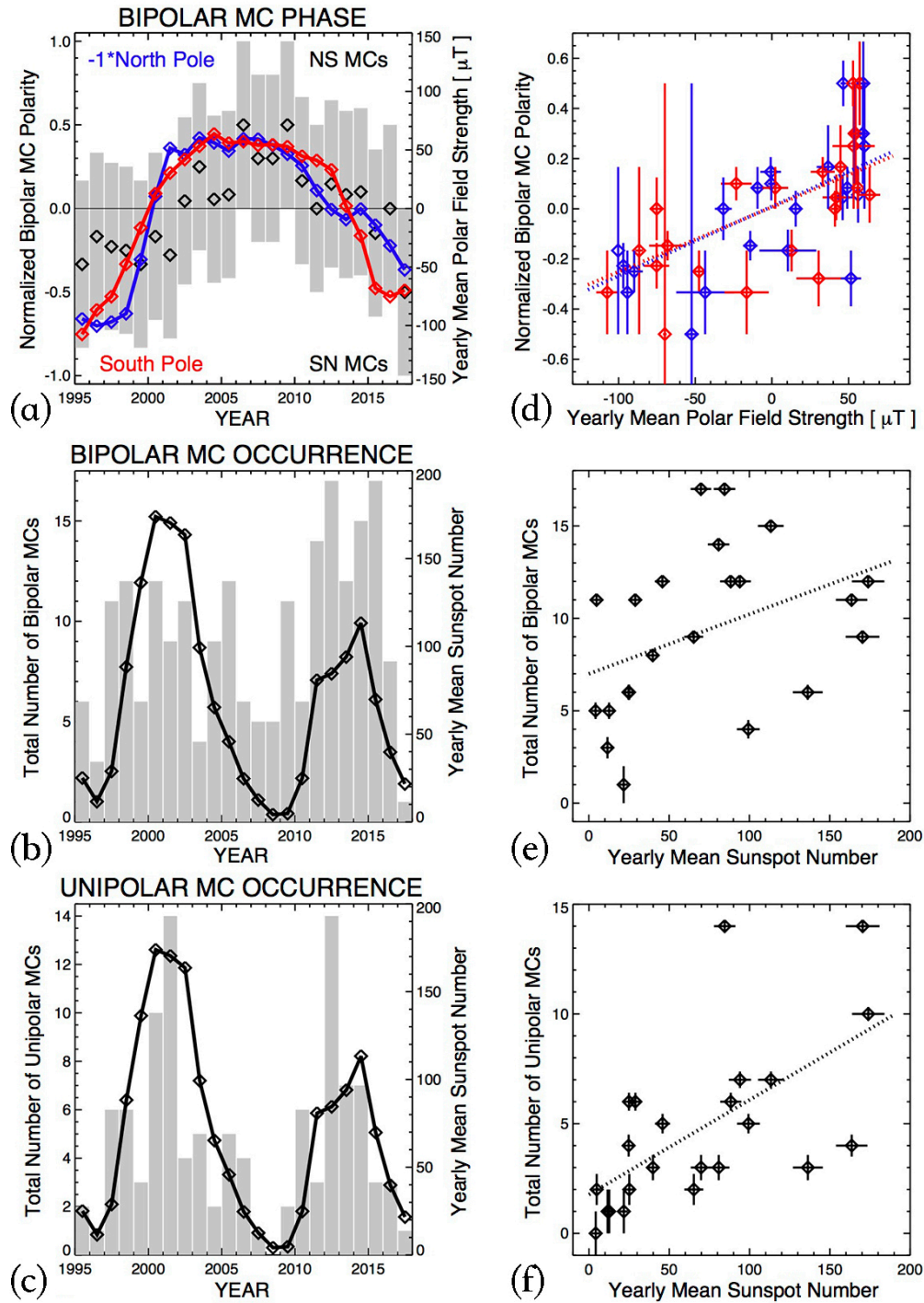

**Figure 10.** Statistic analyses of MC parameters *vs*. solar-cycle index. (**a**) solar polar-field strength over bipolar MC polarity phase, *blue* for north polar field multiplied by -1 and *red* for south polar field, and *black diamonds* are the midpoint of each bars representing normalized annual counts of NS and SN bipolar MCs. (**b**) sunspot numbers over bipolar MC occurrence. (**c**) sunspot numbers over unipolar MC occurrence. (**d**), (**e**) and (**f**) scatter plot and linear fit of the MC parameter and solar-cycle index corresponding to the panel on the left, respectively.

Eighty-five percent of total 303 MCs during the recent two solar cycles contain a south Bz component for some part or for the entire duration of their passage over the Earth, highlighting the space-weather importance of MCs. The MCs with the highest speed, strongest magnetic magnitude, and largest Bz south component occur in the declining phase of the solar cycle from the solar maximum to one or two years before the next solar minimum. When bipolar MCs are grouped by the type of their solar source, the polarity of MCs from active-region flares does not show solar-cycle dependence, while those from large filament sources and uncertain sources show a similar solar-cycle dependence. Based on these new findings, the solar-cycle dependence of the polarity of bipolar MCs is mostly from the contributions of quiescent filament eruptions and some weak events of uncertain sources (including stealth CMEs), and the fluctuations or outliers in the picture are mostly from active-region flare-associated MC events.

Active-region magnetic fields are stronger and can have considerably more complexity, variation, and evolution of their flux distributions and in their orientations with respect to the surrounding corona (Ugarte-Urra, Warren, and Winebarger, 2007). Large, quiescent filaments are from decayed magnetic regions whose on-disk polarity often agrees with the large-scale magnetic field of the corona (Li and Luhmann, 2006). A part of the MC events from uncertain sources are identified as stealth CMEs, which originate from greater heights in the corona and involve the gradual eruption of larger-scale helmet streamer fields. The global helmet streamer belt fields are weaker, less complex, and typically reflect the orientation of the global solar dipole. CMEs from strong, active-region magnetic structures may rotate more after eruptions due to their stronger internal stress (*e.g.* Lynch *et al*., 2009) in addition to their highly variable active-region source and flux-rope orientations (Leamon, Canfield, and Pevtsov, 2002; Leamon *et al*., 2004). Therefore the magnetic-field orientations of MCs from active regions are expected to be less predictable in a general sense. Unfortunately, this group of events poses the most threat to space weather because they are the source-region category that usually produces the most energetic eruptions resulting in the strongest MC magnetic fields and fastest MC speeds. Better understanding of the detailed processes of CME eruptions from different sources and their heliospheric propagation is needed for successful space-weather forecasting.

The source regions of Earth-impacting MCs are located within 45° of solar disk center with a few exceptions. On average, the filament sources tend to be located at higher latitudes than flare sources. We find that the MCs with the highest speeds, largest total *B* magnitudes and sheath Bz south are from source regions closer to the solar disk center, within about 30° latitude and longitude. The velocity and field parameters of MCs with unclear sources are all generally small/weak. While this type of CME may be the most difficult to predict, statistically, these events should not have severe space-weather impacts. Our MC database is inclusive of all the *in-situ* MC events observed at 1 AU in the last 11 years, 2007—2017. The combination of the Bz south fields in the sheath regions and SN MCs are the cause of the largest geomagnetic storms. Also these cases show that storms can be given only by the south Bz in sheath regions. Thus, for space-weather forecasting, the magnetic field in the sheath regions of MCs should also be taken into consideration in addition to the field within MCs.

We updated the record of the solar-cycle dependence of bipolar magnetic clouds (MCs) using newly available solar-wind and IMF data obtained near 1AU and the ecliptic plane. The MC polarity trend, presented in a normalized format to emphasize solar-cycle phase information, now span about four solar cycles (42 years) from 1976 to 2017. The MC polarity on the rising phase of the solar cycle is clearly dominated by one type of bipolar MC, either NS or SN. During a few years around a solar minimum, MCs are always almost pure NS or SN with little exception. Mixed polarities begin to appear with the increase of solar activity and approaching the solar maximum. The predominant MC polarity reverses within the declining phase. The last sunspot minimum and maximum occurred in 2009 and 2014, respectively. The Sun is currently on the declining phase of Cycle 24. The bipolar MC polarity in 2013 to 2017 shows large fluctuations between the occurrences of the two types of bipolar MCs, while maintaining the phase of the cyclic polarity reversal switching from NS MCs as the major type to SN MCs as the major type.

The implications of ICMEs in the solar magnetic cycle have been conceptually considered in the past as part of the picture of the solar dynamo operation (*e.g.* Käpylä, Korpi, and Brandenburg, 2010; Warnecke and Brandenburg, 2010, 2014; Warnecke, Brandenburg, and Mitra, 2011; Warnecke *et al.*, 2012). Although this topic is not the emphasis of this report, our finding that quiescent filament CMEs and stealth CMEs give rise to the MC polarity cycle suggests that these MCs participate in removing the like polarity of the solar dipole field. Those MCs related to flares, on the other hand, likely play a role in removing and relaxing the toroidal fields of solar active regions. In this respect, the phases of the changes in the MC polarities compared to the sunspot-number and polar-field-polarity cycles (see Figure 9) are key observational results. Care must be taken, however, to take into account the observational biases represented by the near-ecliptic observations at 1 AU. Additional polarity analyses performed on *Ulysses* high-latitude ICMEs, as well as the upcoming *Solar Orbiter* ICMEs, should prove interesting and complementary in this regard.

**Acknowledgements**

This work was supported in part by NASA Grant NNX15AG09G to the University of California, Berkeley for the STEREO/IMPACT investigation. Y. Li acknowledges the support of NASA LWS grant NNX15AB80G. Y. Li and B.J. Lynch are supported by NSF grant SHINE-1622495 and NASA HSR grant NNX17AI28G. B.J. Lynch is also supported by the Coronal Global Evolutionary Model (CGEM) project NSF AGS-1321474. We acknowledge ACE/MAG and /SWEPAM level 2 data and OMNI2 data, STEREO/IMPACT and /SECCHI data, SDO/AIA imaging data. We acknowledge LASCO CME catalog generated and maintained at the CDAW Data Center by NASA and The Catholic University of America in cooperation with the Naval Research Laboratory. SOHO is a project of international cooperation between ESA and NASA.

**Disclosure of Potential Conflicts of Interest**
The authors declare that they have no conflicts of interest.

# References

Alzate, N., Morgan, H.: 2017, Identification of Low Coronal Sources of "Stealth" Coronal Mass Ejections Using New Image Processing Techniques, *Astrophys. J.*, **840**, 103, doi: 10.3847/1538-4357/aa6caa

Bothmer, V., Rust, D. M.: 1997, The field configuration of magnetic clouds and the solar cycle, In Crooker, N. Joselyn, J. A., Feynman, J. (eds.) *Coronal Mass Ejections, Geophys. Monogr. Ser.* **99**, AGU, Washington, D. C., 139, doi: 10.1029/GM099p0139

Bothmer, V., Schwenn, R.: 1998, The structure and origin of magnetic clouds in the solar wind, *Ann. Geophys.*, **16**, 1–24, doi: 10.1007/s00585-997-0001-x

Burlaga, L. F.: 1988, Magnetic clouds and force-free fields with constant alpha, *J. Geophys. Res.*, **93**, 7217-7224, doi: 10.1029/JA093iA07p07217

Chi, Y., Shen, C. Wang, Y., Xu, M., Ye, P., Wang, S.: 2016, Statistical Study of the Interplanetary Coronal Mass Ejections from 1996 to 2015, *Solar Phys.*, **291**, 2419-2439, doi: 10.1007/s11207-016-0971-5

D'Huys, E., Seaton, D. B., Poedts, S., Berghmans, D.: 2014, Observational characteristics of coronal mass ejections without low-coronal signatures, *Astrophys. J.*, **795**, 49, doi: 10.1088/0004-637X/795/1/49

Gopalswamy, N., Akiyama, S., Yashiro, S., Michalek, G., Lepping, R. P.: 2008, Solar sources and geospace consequences of interplanetary magnetic clouds observed during solar cycle 23, *J. Atm. Sol.-Terr. Phys.*, **70**, 245-253, doi: 10.1016/j.jastp.2007.08.070

Hess, P., Zhang, J.: 2017, A Study of the Earth-Affecting CMEs of Solar Cycle 24, *Solar Phys.*, **292**, 80, doi: 10.1007/s11207-017-1099-y

Jian, L. K., Russell, C. T., Luhmann, J. G.: 2011, Comparing Solar Minimum 23/24 with Historical Solar Wind Records at 1 AU, *Solar Phys.*, **274**, 321-344, doi: 10.1007/s11207-011-9737-2

Jian, L. K., Russell, C. T., Luhmann, J. G., Skoug, R. M.: 2006, Properties of Interplanetary Coronal Mass Ejections at One AU During 1995—2004, *Solar Phys.*, **239**, 393-436, doi: 10.1007/s11207-006-0133-2

Jian, L. K., Russell, C. T., Luhmann, J. G., Galvin, A. B., Simunac, K. D. C.: 2013, Solar wind observations at STEREO: 2007—2011, in SOLAR WIND 13: Proceedings of the Thirteenth International Solar Wind Conference, **CP-1539**, AIP, Melvill, 195, doi: 10.1063/1.4811021

Käpylä, P. J., Korpi, M. J., Brandenburg, A.: 2010, Open and closed boundaries in large-scale convective dynamos, *Astron. and Astrophys.,* **518**, A22, doi: 10.1051/0004-6361/200913722

Kilpua, E. K. J., Luhmann, J. G., Jian, L. K., Russell, C. T., Li, Y.: 2014, Why have geomagnetic storms been so weak during the recent solar minimum and the rising phase of cycle 24?, *J. Atm. Sol.-Terr. Phys.,* **107**, 12-19 doi: 10.1016/j.jastp.2013.11.001

Krista, L. D., Reinard, A. A.: 2017, Statistical Study of Solar Dimmings Using CoDiT, *Astrophys. J.,* **839**, 50, doi: 10.3847/1538-4357/aa6626

Leamon, R. J., Canfield, R. C., Pevtsov, A. A.: 2002, Properties of magnetic clouds and geomagnetic storms associated with eruption of coronal sigmoids, *J. Geophys. Res.,* **107**, 1234, doi: 10.1029/ 2001JA000313

Leamon, R. J., Canfield, R. C., Jones, S. L., Lambkin, K., Lundberg, B. J., Pevtsov, A. A.: 2004, Helicity of magnetic clouds and their associated active regions, *J. Geophys. Res.,* **109**, A05106, doi: 10.1029/ 2003JA010324

Lee, C. O., Luhmann, J. G., Zhao, X. P., Liu, Y., Riley, P., Arge, C. N., Russell, C. T., de Pater, I.: 2009, Effects of the Weak Polar Fields of Solar Cycle 23: Investigation Using OMNI for the STEREO Mission Period, *Solar Phys.,* **256**, 345-363, doi: 10.1007/s11207-009-9345-6

Li, Y., Luhmann, J. G.: 2006, Coronal Magnetic Field Topology over Filament Channels: Implication for Coronal Mass Ejection Initiations, *Astrophys. J.,* **648**, 732-740, doi: 10.1086/505686

Li, Y., Luhmann, J. G., Lynch, B. J., Kilpua, E. K. J.: 2011, Cyclic reversal of magnetic cloud poloidal field, *Sol. Phys.,* **270**, 331–346, doi: 10.1007/s11207-011-9722-9

Li, Y., Luhmann, J. G., Lynch, B. J., Kilpua, E. K. J.: 2014, Magnetic clouds and origins in STEREO era, *J. Geophys. Res. Space Physics*, **119**, 3237–3246, doi: 10.1002/2013JA019538

Lugaz, N., Temmer, M., Wang, Y., Farrugia, C. J.: 2017, The Interaction of Successive Coronal Mass Ejections: A Review, *Solar Phys.,* **292**, 64, doi: 10.1007/s11207-017-1091-6

Lynch, B. J., Antiochos, S. K., Li, Y., Luhmann, J. G., DeVore, C. R.: 2009, Rotation of coronal mass ejections during eruption, *Astrophys. J.*, **697**, 1918–1927, doi: 10.1088/0004-637X/697/2/1918

Lynch, B. J., Masson, S., Li, Y., DeVore, C. R., Luhmann, J. G., Antiochos, S. K., Fisher, G. H.: 2016, A model for stealth coronal mass ejections, *J. Geophys. Res.,* **121**, 10,677–10,697, doi: 10.1002/2016JA023432

Mulligan, T., Russell, C. T., Luhmann, J. G.: 1998, Solar cycle evolution of the structure of magnetic clouds in the inner heliosphere, *Geophys. Res. Lett.*, **25**, 2959–2962, doi:10.1029/98GL01302

Riley, P.: 2016, Predicting Bz: Baby Steps, in *American Geophysical Union, Fall General Assembly 2016*, abstract #SM23C-01

Riley, P., Love, J. J.: 2017, Extreme geomagnetic storms: Probabilistic forecasts and their uncertainties, *Space Weather*, **15**, 53-64, doi: 10.1002/2016SW001470

Richardson, I. G., Cane, H. V.: 2010, Near-Earth Interplanetary Coronal Mass Ejections During Solar Cycle 23 (1996—2009): Catalog and Summary of Properties, *Solar Phys.*, **264**, 189-237, doi: 10.1007/s11207-010-9568-6

Richardson, I. G., Cane, H. V.: 2013, Solar wind drivers of geomagnetic storms over more than four solar cycles, in SOLAR WIND 13: Proceedings of the Thirteenth International Solar Wind Conference, **CP-1539**, AIP, Melvill, 422, doi: 10.1063/1.4811075

Robbrecht, E., Patsourakos, S., Vourlidas, A.: 2009, No trace left behind: STEREO observation of a coronal mass ejection without low coronal signatures, *Astrophys. J.*, **701**, 283–291, doi:10.1088/0004-637X/701/1/283

Thompson, B. J., Plunkett, S. P., Gurman, J. B., Newmark, J. S., St. Cyr, O. C., Michels, D. J.: 1998, SOHO/EIT observations of an Earth-directed coronal mass ejection on May 12, 1997, *Geophys. Res. Lett.*, **25**, 2465-2468, doi: 10.1029/98GL50429

Ugarte-Urra, I., Warren, H. P., Winebarger, A. W.: 2007, The Magnetic Topology of Coronal Mass Ejection Sources, *Astrophys. J.*, **662**, 1293-1301, doi: 10.1086/514814

Warnecke, J., Brandenburg, A.: 2010, Surface appearance of dynamo-generated large-scale fields, *Astron. and Astrophys.*, **523**, A19, doi: 10.1051/0004-6361/201014287

Warnecke, J., Brandenburg, A.: 2014, Coronal influence on dynamos, in Magnetic Fields throughout Stellar Evolution, *Proceedings of the International Astronomical Union, IAU Symposium*, **302**, 134-137, doi: 10.1017/S1743921314001884

Warnecke, J., Brandenburg, A., Mitra, D.: 2011, Plasmoid ejections driven by dynamo action underneath a spherical surface, in Advances in Plasma Astrophysics, *Proc. Internat. Astron. Union Symp.*, **274**, 306-309, doi: 10.1017/S1743921311007186

Warnecke, J., Käpylä, P. J., Mantere, M. J., Brandenburg, A.: 2012, Coronal ejections from convective spherical shell dynamos, in Comparative Magnetic Minima: Characterizing quiet times in the Sun and Stars, *Proc. Internat. Astron. Union Symp.*, **286**, 154-158, doi: 10.1017/S1743921312004772

Zhang, G., Burlaga, L. F.: 1988, Magnetic clouds, geomagnetic disturbances, and cosmic ray decreases, *J. Geophys. Res.,* **93**, 2511-2518, doi: 10.1029/JA093iA04p02511

Zhang, J., Liemohn, M. W., Kozyra, J. U., Lynch, B. J., Zurbuchen, T. H.: 2004, A statistical study of the geoeffectiveness of magnetic clouds during high solar activity years, *J. Geophys. Res.,* **109**, A09101, doi: 10.1029/2004JA010410

Zhang, J., Richardson, I. G., Webb, D. F., Gopalswamy, N., Huttunen, E., Kasper, J. C., Nitta, N. V., *et al.*: 2007, Solar and interplanetary sources of major geomagnetic storms (Dst <= -100 nT) during 1996—2005, *J. Geophys. Res.,* **112**, A10102, doi: 10.1029/2007JA012321